\documentclass[10pt,conference]{IEEEtran}

\usepackage{epsfig,setspace,amsmath,epsf,amssymb,bm,theorem,cite, graphicx, epstopdf, algorithm, algpseudocode,float,color,accents}
\usepackage{multirow}
\usepackage{authblk}
\usepackage[table,xcdraw]{xcolor}

\newtheorem{lemma}{Lemma}

\IEEEoverridecommandlockouts

\newcommand{\bt}{{\boldsymbol{\tau}}}
\newcommand{\st}{{\text{s.t.}}}

\allowdisplaybreaks

\begin{document}

\title{Timely Private Information Retrieval}

\author[1]{Karim Banawan}
\author[2]{Ahmed Arafa}
\author[3]{Sennur Ulukus}
\affil[1]{\normalsize Electrical Engineering Department, Faculty of Engineering, Alexandria University}
\affil[2]{\normalsize Electrical and Computer Engineering Department, University of North Carolina at Charlotte}
\affil[3]{\normalsize Department of Electrical and Computer Engineering, University of Maryland}

\maketitle

\begin{abstract}
We introduce the problem of \emph{timely} private information retrieval (PIR) from $N$ non-colluding and replicated servers. In this problem, a user desires to retrieve a message out of $M$ messages from the servers, whose contents are continuously updating. The retrieval process should be executed in a timely manner such that no information is leaked about the identity of the message. To assess the timeliness, we use the \emph{age of information} (AoI) metric. Interestingly, the timely PIR problem reduces to an AoI minimization subject to PIR constraints under \emph{asymmetric traffic}. We explicitly characterize the optimal tradeoff between the PIR rate and the AoI metric (peak AoI or average AoI) for the case of $N=2$, $M=3$. Further, we provide some structural insights on the general problem with arbitrary $N$, $M$. \vspace{-.1in}
\end{abstract}

\section{Introduction}
Private information retrieval (PIR), introduced in \cite{PIR_ORI}, investigates the privacy of downloaded content from distributed databases. In classical PIR, a user needs to retrieve a message without disclosing the identity of the desired message to any individual server. Sun and Jafar in \cite{PIR} investigate the PIR problem using information-theoretic measures. The goal of \cite{PIR} is to characterize the PIR capacity, which is the supremum of PIR retrieval rates among all retrieval schemes. The PIR rate is the ratio between the desired message size and the total download from the servers. The optimal scheme in \cite{PIR} is a greedy retrieval scheme that downloads \emph{symmetric} amount of downloaded bits from each server. Several variants of the classical problem were studied afterwards, e.g., \cite{JafarColluding, Staircase_PIR, SPIR, ChaoTian_leakage, MM-PIR, BPIRjournal, tandon2017capacity, wei2017fundamental, PIR_cache_edge, kadhe2017private, chen2017capacity, wei2017capacity, MMPIR_PSI, LiConverse, PrivateComputation, mirmohseni2017private, PrivateSearch, StorageConstrainedPIR, chao-tian, efficient_storage_ITW2019, PIR_decentralized, heteroPIR, Karim_nonreplicated, PIR_WTC_II, arbmsgPIR, securePIRcapacity, securestoragePIR,  XSTPIR, ChaoTian_coded_minsize, KarimAsymmetricPIR, noisyPIR,  PIR_networks,PSIjournal,MP-PSI_journal}. 

All these works, however, do not consider the \emph{timeliness} of the retrieval process. This is crucial in real-time applications. For instance: Investors in the stock market need to retrieve information about the desired stocks without leaking information about their interests. The stock prices change rapidly over time. This calls for retrieving stock information \emph{privately} and \emph{timely}, motivating the \emph{timely PIR} problem. The timeliness can be measured by the \emph{age of information} (AoI) metric, defined as the time elapsed since the latest message has been retrieved. AoI has been originally studied in queuing networks, e.g., \cite{yates_age_1, ephremides_age_random}, and has found its way into other numerous contexts, see, e.g., \cite{modiano-age-bc, sun-age-mdp, jing-age-online, himanshu-age-source-coding, baknina-updt-info, zhou-age-iot, yates-age-mltpl-src, bacinoglu-aoi-eh-finite-gnrl-pnlty, zhang-arafa-aoi-pricing-wiopt, bedewy-aoi-multihop, talak-aoi-delay, arafa-age-online-finite, yang-arafa-aoi-fl, arafa-aoi-estimate-ou, ornee-aoi-estimation-ou}, and the recent survey in \cite{aoi-JSAC-survey}.

From a technical standpoint, the timely PIR problem investigates an interesting tension between privacy and AoI. Specifically, minimizing AoI (if privacy is ignored) necessitates downloading from a \emph{single} server with minimal delay statistics. This however results in the worst achievable PIR rate. This extreme case poses an interesting question: {\bf what is the optimal tradeoff between the PIR rate and the AoI?}   

\begin{figure}[t]
\centering
\includegraphics[width=0.4\textwidth]{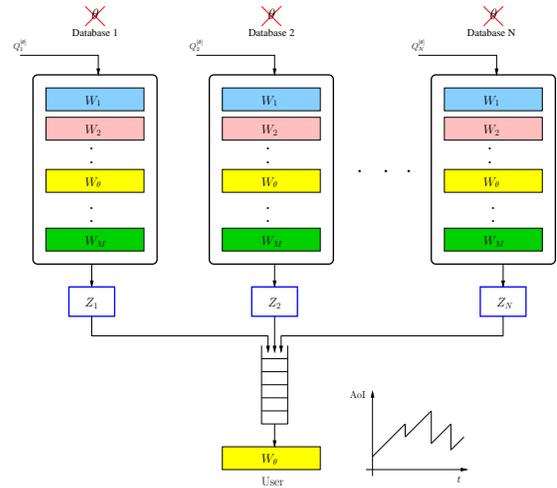}
\vspace{-.1in}
\caption{The timely PIR problem.}
\label{fig-sys_mod}
\vspace{-0.22in}
\end{figure}


In this paper, we introduce the timely PIR problem. A user aims at correct retrieval of a message $W_\theta$ (of size $L$) out of $M$ messages from $N$ non-colluding and replicated servers with different delay statistics. To that end, the user designs queries to minimize the AoI metric such that the queries do not leak $\theta$ to any individual server and achieve a minimum PIR rate of $R_{\min}$. This reduces to optimizing the download size from each server. Interestingly, the asymmetry of the download sizes fundamentally links the timely PIR problem to the PIR problem with asymmetric traffic constraints \cite{KarimAsymmetricPIR}, which results in $N!\binom{N+M-1}{M}$ constraints on the achievable PIR rate. We consider two age metrics, namely, the peak-age metric and the average-age metric. For the case of $N=2$ servers and $M=3$ messages, we analytically characterize the optimal tradeoff between the minimum PIR rate $R_{\min}$ and AoI. The peak AoI problem is expressed as a linear program, while an inner-outer minimization procedure is used to solve the average AoI problem. In both cases, the optimal age-metric is a non-decreasing function of $R_{\min}$. For the general problem with an arbitrary number of servers and messages, we show that 1) for peak AoI, a user downloads more bits from servers with lower expected delay, 2) if delay statistics are identical, there is a synergy between optimizing PIR rate and AoI.    

\section{The Timely PIR Problem}

Consider a dynamic distributed storage setting with $N$ replicated and non-colluding servers. Each server possesses a library of $M$ i.i.d. messages of size $L$ bits. We denote these messages by $\{W_m\}_{m=1}^M$. A user is interested in retrieving one of these messages, $W_\theta$, without revealing the identity of the status update $\theta$ to any individual server. Following a completed retrieval process, the contents of the servers may change to a new set of messages, in an i.i.d. manner. In order to follow this changing nature, the user aims at retrieving its intended message in a {\it timely} manner while preserving privacy. Thus, a {\it status update} is received with each successful retrieval process.


Towards retrieving $W_\theta$, the user submits a query $Q_n^{[\theta]}$ to server $n\in[N]$. Server $n$ then truthfully responds with an answer string $A_n^{[\theta]}$, which is a deterministic function of the query $Q_n^{[\theta]}$ and the contents of the servers $\{W_m\}_{m=1}^M$, i.e.,
\begin{align}
    H\left(A_n^{[\theta]}\Big|Q_n^{[\theta]},W_{1:M}\right)=0, \quad n\in[N].
\end{align}
In this work, we assume that the user incurs a negligible delay to convey the queries to the servers.\footnote{We focus on the case in which the sole source of the delay is in the downlink---while the servers are responding to the user. This assumption can be motivated by the fact that the upload cost in PIR problems does not scale with the message size $L$ in contrast to the download cost \cite{PIR}. Moreover, the queries can be reused if the user is interested in the same message.} The returned answer strings are time-stamped by the server at the exact time instant of receiving the query.\footnote{The servers' time stamps are the same since the upload delay is negligible.} The user incurs a random delay, $Z_{n,i}$ time units, in order to the receive the $i$th answer from server $n$. This models the total delay of the server including the processing and transmission delays, in addition to the propagation delay to the user. The random variables $\{Z_{n,i}\}_{i=1}^\infty$ are i.i.d. with mean $\mu_n$ and variance $\sigma^2_n$. The statistics of the server delays are known to the user.

The user designs queries that result in downloading $d_n$ bits from server $n$. The value of $d_n$ is fixed across different status updates, i.e., the same number of bits $d_n$ are downloaded from server $n$ whenever a new status update is sought. The answer strings from all servers are received through a single shared queue (see Fig.~\ref{fig-sys_mod}), and therefore the $j$th status update requires
\begin{align} \label{eq_delay-model}
    T_j\triangleq\sum_{n=1}^N\sum_{i=(j-1)d_n+1}^{jd_n} Z_{n,i}
\end{align}
time units to be received in full. We denote by an {\it epoch} the time elapsed in between two consecutive status updates. Since $\{Z_{n,i}\}_{i=1}^\infty$ are i.i.d. and $\{d_n\}_{n=1}^N$ are fixed, it follows that the epoch times $\{T_j\}_{j=1}^\infty$ are i.i.d. $\sim T$. We assume that the servers do not update their message contents within the same epoch.


Furthermore, the user designs the queries such that they satisfy the following PIR constraints:\\
\noindent {\bf Correctness.} The user should be able to reconstruct the message with no error given the answer strings, i.e.,
\begin{align}\label{correctness}
    H\left(W_\theta\Big|A_{1:N}^{[\theta]},Q_{1:N}^{[\theta]}\right)=0.
\end{align}
\noindent {\bf Perfect Privacy.} The queries should not leak any information about the identity of the message, i.e.,
\begin{align}\label{privacy}
    I\left(\theta; Q_n^{[\theta]}\right)=0, \quad n\in[N]. 
\end{align}
\noindent {\bf Minimum Retrieval Rate.} We use the retrieval rate as the performance metric of the retrieval scheme. The retrieval rate, $R$, is the ratio between the status update size $L$ and the \emph{expected}\footnote{By measuring the download cost in the expected sense, we effectively allow for \emph{mixed} retrieval strategies, i.e., we allow stochastic time sharing (across epochs) between pure retrieval strategies in this work.} total download cost $D$, i.e., $R=\frac{L}{\mathbb{E}[D]}$. The designed queries need to achieve a minimum retrieval rate $R_{\min}$,
\begin{align}\label{rate constraint}
    R \geq  R_{\min} \quad \Leftrightarrow \quad \mathbb{E}[D] \leq D_{\max}=\frac{L}{R_{\min}}. 
\end{align}


We use the AoI metric to assess the timeliness of the received status updates. The AoI at time $t$ is defined as
\begin{align}
    a(t)\triangleq t-u(t),
\end{align}
where $u(t)$ denotes the time stamp of the latest received status update before time $t$. We focus on minimizing two AoI metrics. The first is the long-term {\it peak} average AoI:
\begin{align} \label{eq_pAoI}
    \texttt{pAoI}\triangleq\limsup_{T\rightarrow\infty}\mathbb{E}\left[\frac{1}{l(T)}\sum_{j=1}^{l(T)}a\left(s_j^-\right)\right],
\end{align}
where $l(T)$ denotes the number of status updates received by time $T$, and $s_j^-$ denotes the time right before receiving the $j$th status update. The second is the long-term {\it time} average AoI:
\begin{align} \label{eq_aAoI}
    \texttt{aAoI}\triangleq\limsup_{T\rightarrow\infty}\frac{1}{T}\mathbb{E}\left[\int_0^Ta(t)dt\right],
\end{align}
The expectations in (\ref{eq_aAoI}) and (\ref{eq_pAoI}) are taken over the distributions of the incurred delays, as manifested in (\ref{eq_delay-model}). We focus on {\it zero-wait} policies in which a new status update is requested right after the previous one is received.\footnote{Zero-wait policies are generally optimal when minimizing \texttt{pAoI} but may be suboptimal when minimizing \texttt{aAoI} \cite{sun-age-mdp}. However, since the purpose of this work is to introduce the notion of timely PIR, we focus on zero-wait policies for simplicity of presentation.} 

\begin{figure}[t]
\centering
\includegraphics[width=0.22\textwidth]{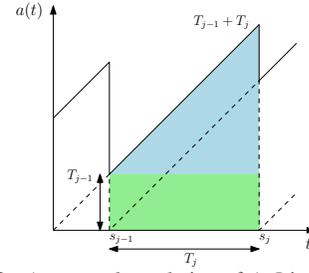}
\vspace{-.15in}
\caption{An example evolution of AoI in epoch $j$.}
\label{fig_aoi-evol-ex}
\vspace{-.2in}
\end{figure}

In Fig.~\ref{fig_aoi-evol-ex}, we show an example of how the AoI may evolve during epoch $j$. From the figure, one can see that the epoch's peak AoI has a value of $T_{j-1}+T_j$. Since $T_j$'s are i.i.d., one can show that (\ref{eq_pAoI}) reduces to
\begin{align}
    \texttt{pAoI}=\mathbb{E}\left[T_{j-1}+T_j\right]=2\mathbb{E}\left[T\right].
\end{align}
Similarly, one can compute the area of epoch $j$ and use the renewal-reward theorm \cite{ross1996stochastic} to show that (\ref{eq_aAoI}) reduces to
\begin{align}
    \texttt{aAoI}\!=\!\!\frac{1}{\mathbb{E}\!\left[T_j\!\right]}\!\!\left[\mathbb{E}\left[T_{j-1}T_j\right]\!+\!\frac{1}{2}\mathbb{E}\left[T_j^2\!\right]\!\right]\!=\!\mathbb{E}\left[T\right]\!+\!\frac{\mathbb{E}\left[T^2\right]}{2\mathbb{E}\left[T\right]}.
\end{align}

We now formulate the {\bf timely PIR problem} as follows:
\begin{align}\label{general problem}
    \min_{\{d_n,~Q_n^{\theta}\}_{n=1}^N} &\quad  \text{\texttt{pAoI} or \texttt{aAoI}} \notag\\
    \st \hspace{.2in}  &\quad H\left(A_n^{[\theta]}\Big|Q_n^{[\theta]},W_{1:M}\right)=0, \quad n\in[N] \notag\\
          &\quad H\left(W_\theta\Big|A_{1:N}^{[\theta]},Q_{1:N}^{[\theta]}\right)=0 \notag\\
          &\quad I\left(\theta; Q_n^{[\theta]}\right)=0, \quad\quad\quad\quad\quad\: n\in[N] \notag\\
          &\quad R \geq R_{\min}.
\end{align}
Observe that designing the queries $\{Q_n^\theta\}_{n=1}^N$ implicitly leads to the number of downloads from each server $\{d_n\}_{n=1}^N$ and the expected download cost $\mathbb{E}\left[D\right]$. 

Problem (\ref{general problem}) holds an interesting tension. For instance, if $R_{\min}$ allows it, the user may opt to retrieve its message from the server with the {\it most favorable statistics} (AoI-wise). However, this would come at the expense of {\it downloading more bits} to satisfy privacy. We shall see that it may be more rewarding to interact with servers with {\it individually} less favorable statistics so as to maintain privacy with a fewer number of downloads, and achieve an {\it overall} smaller AoI.

For a given server statistic vectors $\boldsymbol{\mu}\triangleq[\mu_1 \:\: \mu_2 \:\: \cdots \:\: \mu_N]^T$ and $\boldsymbol{\sigma}\triangleq[\sigma_1^2 \:\: \sigma_2^2 \:\: \cdots \:\: \sigma_N^2]^T$, we aim at characterizing the optimal tradeoff between the long-term average AoI and the retrieval rate. More specifically, we aim at characterizing $\alpha^*(R_{\min})$, where $\alpha^*(\cdot)$ is the optimal value function of the optimization problem in \eqref{general problem} for all $\frac{1}{M} \leq R_{\min} \leq C_{\text{PIR}}$.\footnote{The achievable PIR rate is bounded below by $\frac{1}{M}$, which corresponds to the trivial scheme of downloading all messages from one server, and bounded above by the PIR capacity $C_{\text{PIR}}=\left(1+\frac{1}{N}+\cdots+\frac{1}{N^{M-1}}\right)^{-1}$ \cite{PIR}.}

\section{Problem Re-Formulation:\\PIR with Asymmetric Traffic}

In this section, we focus on disentangling the constraint set of problem \eqref{general problem}. With a slight abuse of notation, we refer to the expected download cost by merely $D$. We will also allow for non-integer solutions of the download size vector $\mathbf{d}\triangleq[d_1 \quad d_2 \quad \cdots \quad d_N]^T$, and overcome this by stochastic time sharing between well-defined PIR schemes. Time sharing will be shown optimal for the case of minimizing $\texttt{pAoI}$, and only slightly suboptimal for the case of minimizing $\texttt{aAoI}$.

We now express the AoI in terms of $\boldsymbol{\mu}$, $\boldsymbol{\sigma}$ and $\mathbf{d}$. Given the model in (\ref{eq_delay-model}), and that epochs are i.i.d., one can write
\begin{align}
    \mathbb{E}\left[T\right]=&\sum_{n=1}^N\sum_{i=1}^{d_n}\mathbb{E}\left[Z_{n,i}\right]=\sum_{n=1}^N \mu_n d_n=\boldsymbol{\mu}^T \mathbf{d}, \\
    \mathbb{E}\left[T^2\right]&=\sum_{n=1}^N \sigma_n^2 d_n +(\boldsymbol{\mu}^T \mathbf{d})^2=\boldsymbol{\sigma}^T \mathbf{d}+(\boldsymbol{\mu}^T \mathbf{d})^2.
\end{align}
Based on this, we now have
\begin{align}\label{aAoI}
    \texttt{pAoI}=2\boldsymbol{\mu}^T \mathbf{d}, \quad \texttt{aAoI}=\frac{3}{2}\boldsymbol{\mu}^T\mathbf{d}+\frac{1}{2}\frac{\boldsymbol{\sigma}^T \mathbf{d}}{\boldsymbol{\mu}^T\mathbf{d}}.
\end{align}
Observe that the AoI is solely controlled by the download size vector $\mathbf{d}$. Such vector must satisfy the constraint
\begin{align}
    \mathbf{1}^T \mathbf{d}=D.
\end{align}


Next, we focus on the PIR constraints. Since the user may download more data from one server relative to others, it becomes natural to consider PIR schemes with asymmetric traffic. The work in \cite{KarimAsymmetricPIR} characterizes the optimal retrieval rate $C(\boldsymbol{\tau})$ for an arbitrary server traffic ratio vector $\boldsymbol{\tau}=[\tau_1 \quad \tau_2 \quad \cdots \quad \tau_N]^T$, where $\tau_n=\frac{d_n}{D}$. Reference \cite{KarimAsymmetricPIR} provides an optimal retrieval scheme that satisfies the correctness and privacy constraints for monotonically decreasing traffic vector $\boldsymbol{\tau}$, such that $\tau_1 \geq \tau_2 \geq \cdots \geq \tau_N$ for the cases of $M=2,3$. The results of \cite{KarimAsymmetricPIR} imply that there exist $\binom{M+N-1}{M}$ corner points corresponding to explicit achievable schemes. For any other traffic ratio vector, the optimal scheme is realized by time sharing between adjacent corner points. We focus in this work on the cases of $M=2, 3$, for which the PIR capacity $C(\boldsymbol{\tau})$ is characterized for each $\boldsymbol{\tau}$ as follows \cite{KarimAsymmetricPIR}:
\begin{align}\label{capacityM32}
	C(\bt)\! =\!\!
	\left\{
	\begin{array}{ll}
	\!\!\!\!\displaystyle\min_{n_0 \in [N]}\!\!\! \frac{1+\frac{\sum_{n=n_0+1}^{N} \tau_n}{n_0}}{1+\frac{1}{n_0}}, & M=2 \\
	\!\!\!\!\displaystyle\min_{n_0 \leq n_1 \in [N]}\!\!\!\!\!\!\! \frac{1+\frac{\sum_{n=n_0+1}^{N} \tau_n}{n_0}+\frac{\sum_{n=n_1+1}^{N} \tau_n}{n_0n_1}}{1+\frac{1}{n_0}+\frac{1}{n_0n_1}}, & M=3
	\end{array}
	\right.
\end{align}

Thus, the query design in addition to the PIR constraints in \eqref{general problem} reduce to choosing one of the corner points of \cite{KarimAsymmetricPIR} (or a convex mixture of them). Consequently, the general problem in \eqref{general problem} can be re-formulated by incorporating the capacity expressions for $M=3$ in \eqref{capacityM32},
\begin{align}\label{opt_asymmetric-traffic-problem}
    \min_{\mathbf{d},D} &\quad  \text{\texttt{pAoI} or \texttt{aAoI}} \notag\\
    \st    &\quad 
          \mathbf{d}\geq 0, \quad \mathbf{1}^T \mathbf{d}=D, \quad D \leq D_{\max}\notag\\
          &\quad \min_{\Pi([N])}\left\{ \sum_{n=n_0+1}^N \!\!\!\!\!\!d_{i_n} \!+\!\!\!\!\!\!\sum_{n=n_1+1}^N \!\!\!\!\!\!d_{i_n}\!\!\right\} \geq L\left(\!1\!+\!\frac{1}{n_0}\!+\!\frac{1}{n_0 n_1}\!\!\right)\!-\!D,\: \notag\\ 
          &\qquad \qquad \qquad \qquad  \qquad \qquad \qquad  n_0 \!\leq\! n_1\! \in [N],
\end{align}
where $\Pi([N])$,  is the set of all permutations of the index set $\{i_1, i_2, \cdots, i_N\}$ of the vector $[d_{i_1} \: d_{i_2} \: \cdots \: d_{i_N}]$. We include all the permutations as the optimal ordering is unknown in general. Note that for $M=2$, the third constraint is replaced by $\min_{\Pi([N])}\left\{ \sum_{n=n_0+1}^N \!\!d_{i_n} \right\} \! \geq L\left(1\!+\!\frac{1}{n_0}\!\right)\!-\!D,\:\: n_0 \in [N]$. We focus on problem (\ref{opt_asymmetric-traffic-problem}) in what follows.

\section{Timely PIR for $N=2$ and $M=3$}

In this section, we solve problem \eqref{opt_asymmetric-traffic-problem} explicitly for the special case of $N=2$ servers and $M=3$ messages. In this case, we have $4$ corner points corresponding to $4$ different achievable schemes for PIR under asymmetric traffic constraints (see \cite[Section~V.A]{KarimAsymmetricPIR}). Note that the achievable schemes in \cite[Section~V.A]{KarimAsymmetricPIR} are constructed for different message sizes; specifically, $L$ can be $1,2,4,$ or $8$ for $N=2$ and $M=3$. Since we formulate our problem for a fixed $L$, we choose it to be the least common multiple of all possible message sizes, i.e., we set $L=8$ bits. This implies that the retrieval tables in \cite[Section~V.A]{KarimAsymmetricPIR} need to be repeated to match the chosen message size.

Furthermore, for $N=2$ and $M=3$, the third constraint in \eqref{opt_asymmetric-traffic-problem} can be explicitly written as
\begin{align}\label{explicit constraints}
    d_2\geq \frac{3}{2}L-\frac{1}{2}D,~d_2\geq \frac{5}{2}L-D,~D \geq \frac{7}{4} L.
\end{align}
Note that these constraints are written with the assumption that $d_1 \geq d_2$. For a general $\boldsymbol{\mu}$, and $\boldsymbol{\sigma}$, this may not be optimal. Thus, we need to add two more constraints analogous to \eqref{explicit constraints} after replacing $d_2$ by $d_1$, to cover the case $d_2 \geq d_1$.

\subsection{Peak-Age Minimization Under Perfect Privacy}

Problem (\ref{opt_asymmetric-traffic-problem}) with the $\text{\texttt{pAoI}}$ metric is now given by
\begin{align}\label{peak_2server}
    \min_{\mathbf{d},D} &\quad  2\boldsymbol{\mu}^T \mathbf{d} \notag\\
    \st    &\quad 
          \mathbf{d}\geq 0,~\mathbf{1}^T \mathbf{d}=D,~(7/4)L \leq D \leq L/R_{\min},~L=8 \notag\\
          &\quad d_n \geq \max\left\{\frac{3}{2}L-\frac{1}{2}D,\frac{5}{2}L-D\right\}, \:\: n=1,2. 
\end{align}

Without loss of generality (WLOG), assume that $\mu_1 \leq \mu_2$. This implies that $d_1 \geq d_2$, since the constraint set is symmetric (cf. Lemma~\ref{lemma:order peak}). The optimization problem \eqref{peak_2server} is a linear program (LP), whose solution resides at one of the corner points of the constraint set \cite{boyd2004convex}, which we categorize next.

\subsubsection{$R_{\min} \in [\frac{1}{2},\frac{4}{7}]$}

In this case, we have 2 feasible corner points. The first is the corner point corresponding to the scheme \cite[Table~III]{KarimAsymmetricPIR} with $\tau_2=\frac{3}{7}$, i.e., with $d_1=8$, and $d_2=6$.\footnote{Details of the PIR scheme tables in \cite{KarimAsymmetricPIR} are omitted due to space limits.} This gives $\texttt{pAoI}(R_{\min})\triangleq2\boldsymbol{\mu}^T \mathbf{d}=16\mu_1+12\mu_2$.

The second corner point results from the intersection of the constraints: $D \leq \frac{L}{R_{\min}}$ and $d_2 \geq \frac{5}{2}L-D$. This gives 
\begin{align}
    d_2&=\frac{5}{2}L-\frac{L}{R_{\min}}=\left(\frac{5}{2}-\frac{1}{R_{\min}}\right)L,\\
    d_1&=D-d_2=\left(\frac{2}{R_{\min}}-\frac{5}{2}\right)L.
\end{align}
To achieve such $d_1, d_2$, we employ stochastic time sharing between schemes \cite[Table~III]{KarimAsymmetricPIR} and \cite[Table~IV]{KarimAsymmetricPIR}, i.e., at the start of each epoch, the user randomly applies the scheme \cite[Table~III]{KarimAsymmetricPIR} with probability $\theta=\frac{4}{R_{\min}}-7$ and the scheme \cite[Table~IV]{KarimAsymmetricPIR} with probability $1-\theta$. This gives $\texttt{pAoI}(R_{\min})=16\left[\frac{5}{2}(\mu_2-\mu_2)+\frac{1}{R_{\min}}(2\mu_1-\mu_2)\right]$.
\sloppy Consequently, for $R_{\min} \in [\frac{1}{2},\frac{4}{7}]$, the solution of problem (\ref{peak_2server}) is
\begin{align}
   \alpha^*(R_{\min})\!=\!\min&\bigg\{16\mu_1+12\mu_2, \notag\\ &16\left[\frac{5}{2}(\mu_2\!-\!\mu_2)\!+\!\frac{1}{R_{\min}}(2\mu_1\!-\!\mu_2)\right]\bigg\}.
\end{align}

We note that the scheme \cite[Table~II]{KarimAsymmetricPIR} is strictly sub-optimal since it achieves the same PIR rate of \cite[Table~III]{KarimAsymmetricPIR}, which is $\frac{4}{7}$, while incurring stricltly higher $\texttt{pAoI}$.

\subsubsection{$R_{\min} \in [\frac{1}{3},\frac{1}{2}]$}

In this case, in addition to the corner point of the scheme \cite[Table~III]{KarimAsymmetricPIR}, we have two more corner points. The first is that corresponding to the scheme \cite[Table~IV]{KarimAsymmetricPIR} with $\tau_2=\frac{1}{4}$, i.e., with $d_1=12$, and $d_2=4$. This gives $\texttt{pAoI}(R_{\min})=24\mu_1+8\mu_2$. 

The second corner point results from the intersection of the constraints: $D \leq \frac{L}{R_{\min}}$ and $d_2 \geq \frac{3}{2}L-\frac{1}{2}D$. This gives
\begin{align}
    d_2&=\frac{3}{2}L-\frac{1}{2}\frac{L}{R_{\min}}=\left(\frac{3}{2}-\frac{1}{2R_{\min}}\right)L,\\
    d_1&=D-d_2=\left(\frac{3}{2R_{\min}}-\frac{3}{2}\right)L.
\end{align}
This is achieved by stochastic time sharing between the scheme \cite[Table~I]{KarimAsymmetricPIR} with probability $\theta=\frac{1}{R_{\min}}-2$ and the scheme \cite[Table~IV]{KarimAsymmetricPIR} with probability $(1-\theta)$. This gives $\texttt{pAoI}(R_{\min})=8\left[3(\mu_2-\mu_1)+\frac{1}{R_{\min}}(3\mu_1-\mu_2)\right]$. Consequently, for $R_{\min} \in [\frac{1}{3},\frac{1}{2}]$, the solution of problem (\ref{peak_2server}) is
\begin{align}
   \alpha^*(R_{\min})=\min&\bigg\{16\mu_1+12\mu_2, 24\mu_1+8\mu_2, \notag\\ &8\left[3(\mu_2\!-\!\mu_1)\!+\!\frac{1}{R_{\min}}(3\mu_1\!-\!\mu_2)\right]\bigg\}.
\end{align}

\subsection{Average-Age Minimization Under Perfect Privacy}

The $\texttt{aAoI}$ expression in \eqref{aAoI} has a linear fractional term. To deal with this, we use the Charnes-Cooper transformation in \cite{Charnes-Cooper}. More specifically, we define $t\triangleq\frac{1}{\boldsymbol{\mu}^T \mathbf{d}}$, and change the optimization variable to $\mathbf{x}\triangleq\mathbf{d}\cdot t$. Therefore, we now have
\begin{align}
    \texttt{aAoI}=\frac{3}{2}\frac{1}{t} +\frac{1}{2} \boldsymbol{\sigma}^T \mathbf{x}
\end{align}
We note that this is a convex function in $(\boldsymbol{x},t)$ \cite{boyd2004convex}. The constraint set after transformation becomes
\begin{align}\label{eq_constraint-set-aAoI}
    \!\!\!\mathcal{X}\!\triangleq\!\bigg\{\!\!\left(\mathbf{x},t,D\right)\!:&\mathbf{x}\geq 0, \mathbf{1}^T \mathbf{x}\!=\!Dt, \boldsymbol{\mu}^T \mathbf{x}\!=\!1, x_n \!\geq \! \frac{3}{2}Lt\!-\!\frac{1}{2}Dt,\notag\\
    &x_n \geq \frac{5}{2}Lt-Dt, \frac{7}{4}L \leq D \leq \frac{L}{R_{\min}}\bigg\}.
\end{align}
The constraint set is convex for fixed $D$. This suggests using inner-outer minimization techniques, in which the inner minimization is over $(\mathbf{x},t)$ for a fixed $D$, and the outer minimization is over $D$. The inner problem is a convex problem, while the outer problem can be solved by a one-dimensional line search over $\frac{7}{4}L \leq D \leq \frac{L}{R_{\min}}$.

For the inner minimization, we first note that since $R_{\min}\geq\frac{1}{3}$ holds in our case, the largest value that $D$ takes is $3L$, and hence the PIR constraint $x_n\geq\frac{3}{2}Lt-\frac{1}{2}Dt$ implies $\mathbf{x}\geq0$ and makes it redundant. As in the $\texttt{pAoI}$ case, we assume $\mu_1\leq\mu_2$. Next, we consider the two equality constraints in $\mathcal{X}$. These yield a unique solution for $\mathbf{x}$ provided that $\mu_1\neq\mu_2$, which is given as follows:
\begin{align}
    x_1^*=\frac{\mu_2Dt-1}{\mu_2-\mu_1},\quad x_2^*=\frac{1-\mu_1Dt}{\mu_2-\mu_1}.
\end{align}
Substituting the above in the objective function, taking derivative with respect to $t$ and equating it to $0$ gives
\begin{align} \label{eq_t-star}
    t^*(D)=\sqrt{3\frac{\mu_2-\mu_1}{\sigma_1^2\mu_2-\sigma_2^2\mu_1}\frac{1}{D}}.
\end{align}
However, for this approach to be valid, we must have $t^*(D)\in\mathbb{R}_{++}$ above. Assuming this is the case, it now remains to find the optimal $D^*$ while satisfying the PIR constraints. Towards that end, we substitute $t^*(D)$ in both the objective function and the constraints, and solve the following outer minimization:
\begin{align} \label{opt_outer-min}
    \min_D \quad &\sqrt{3\frac{\sigma_1^2\mu_2-\sigma_2^2\mu_1}{\mu_2-\mu_1}D}+\frac{1}{2}\frac{\sigma_2^2-\sigma_1^2}{\mu_2-\mu_1} \nonumber \\
    \st \quad &\min\left\{\mu_2D-\frac{1}{t^*(D)},\frac{1}{t^*(D)}-\mu_1D\right\} \nonumber \\
    &\hspace{.75in}\geq(\mu_2-\mu_1)\max\left\{\frac{3}{2}L-\frac{1}{2}D,\frac{5}{2}L-D\right\} \nonumber \\
    &\frac{7}{4}L\leq D\leq\frac{L}{R_{\min}},~L=8.
\end{align}
The above shows that $\texttt{aAoI}$ scales with $\sqrt{D}$ (note that $\sigma_1^2\mu_2>\sigma_2^2\mu_1>0$; otherwise $t^*(D)\notin\mathbb{R}_{++}$). The solution of the outer minimization, $D^*$, is given by the least value of $D$ satisfying the constraint set of problem (\ref{opt_outer-min}). Therefore,
\begin{align}
    \alpha^*(R_{\min})=\sqrt{3\frac{\sigma_1^2\mu_2-\sigma_2^2\mu_1}{\mu_2-\mu_1}D^*}+\frac{1}{2}\frac{\sigma_2^2-\sigma_1^2}{\mu_2-\mu_1}.
\end{align}

We now consider the case in which $\mu_1=\mu_2\triangleq\mu$. In this case, the equality constraints in $\mathcal{X}$ do not yield a unique solution for $\mathbf{x}$; they would instead impose that $t^*(D)=\frac{1}{\mu D}$. Assuming WLOG that $\sigma_1^2\leq\sigma_2^2$, further manipulations would lead to the following outer problem (derivation details are omitted due to space limits):
\begin{align}
   \min_D \quad &\frac{3}{2} \mu D\!+\!\frac{\sigma_2^2\!-\!\sigma_1^2}{\mu} \max \left\{\frac{3}{2}\frac{L}{D}\!-\!\frac{1}{2}, \frac{5}{2} \frac{L}{D}\!-\!1\right\}\!+\!\frac{\sigma_1^2}{\mu} \nonumber \\
   \st \quad &\frac{7}{4}L \leq D \leq \frac{L}{R_{\min}},~L=8.
\end{align}
which is a convex problem in $D$ that can be solved, e.g., by a bisection search to give $\alpha^*(R_{\min})$.

Finally, if the value of $t^*$ in (\ref{eq_t-star}) does not yield a positive real solution, or if problem (\ref{opt_outer-min}) is not feasible, then this means that the two servers cannot be simultaneously active. This would require $R_{\min}=\frac{1}{3}$, in which case $d_{n^*}=D=3L$, where $n^*\triangleq\arg\min\frac{9}{2} \mu_n L+\frac{1}{2}\frac{\sigma_n^2}{\mu_n}$, and
\begin{align}
    \alpha^*(R_{\min})=\frac{9}{2} \mu_{n^*} L+\frac{1}{2}\frac{\sigma_{n^*}^2}{\mu_{n^*}}.
\end{align}

\section{Extending to Arbitrary Number of Servers}

For the case with an arbitrary number of servers, we observe that the number of PIR constraints grows as $N ! \binom{N+M-1}{M}$ to encapsulate all possible permutations of the servers. This deems an analytic resolution of the problem extremely challenging. Nevertheless, we have the following structural results (proofs are omitted due to space limits):

\begin{lemma}\label{lemma:order peak}
In problem (\ref{opt_asymmetric-traffic-problem}) with \texttt{pAoI}, if the mean delays are increasing, $\mu_1 \leq \mu_2 \leq \cdots \leq \mu_N$, then the download sizes are decreasing, $d_1 \geq d_2 \geq \cdots \geq d_N$.
\end{lemma}




\begin{lemma} \label{thm_synergy}
In problem (\ref{opt_asymmetric-traffic-problem}) with \texttt{pAoI} or \texttt{aAoI}, if $\mu_n=\mu,~\sigma_n^2=\sigma^2,~\forall n\in[N]$, then $d_n=d=\frac{L}{N C_{PIR}},~\forall n\in[N]$ and the symmetric Sun-Jafar PIR scheme in \cite{PIR} is optimal.

\end{lemma}

 

Lemma~\ref{thm_synergy} shows that for symmetric server statistics, load balancing, which is known to maximize the PIR rate is optimal. Consequently, {\it symmetric statistics produces a synergy between maximizing the PIR rate and minimizing AoI.}

\begin{figure}[t]
	\centering
	\includegraphics[width=0.42\textwidth]{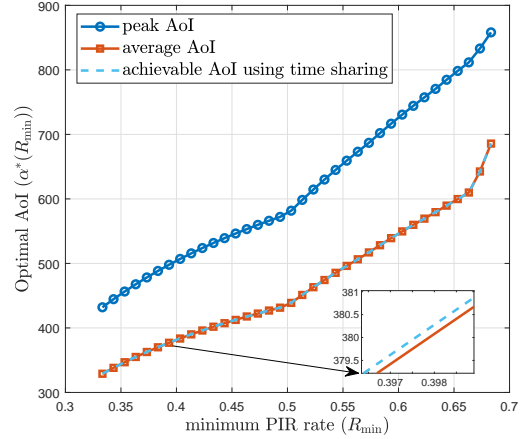}
	\caption{Optimal tradeoff between AoI and the PIR rate for $N=3$, $M=3$, $L=72$, with the servers having $\boldsymbol{\mu}=[1 \: 5 \: 10]$ and $\boldsymbol{\sigma}=[10 \: 5 \: 1]$}
	\label{D_L_curve}
	\vspace{-.2in}
\end{figure}

In Fig.~\ref{D_L_curve}, we show how the optimal AoI $\alpha^*(R_{\min})$ behaves with the minimum PIR rate $R_{\min}$ for the case of $N=3$ servers, $M=3$ messages, and $L=72$ bits. One can observe a clear tradeoff between AoI and the PIR rate, and that time sharing for \texttt{aAoI} has a negligible performance loss.

\section{Conclusions}
In this paper, we introduced the timely PIR problem, in which a user wishes to privately retrieve a message in timely fashion from $N$ replicated and non-colluding servers, whose $M$ messages are continuously updating. We showed that the query design problem in this case reduces to choosing the download sizes from each server, tying the problem to PIR under asymmetric traffic. The optimal tradeoff between the PIR rate and (peak and average) AoI has then been studied.


Several extensions can be pursued. First, reducing the number of constraints by identifying the optimal ordering of the servers and/or the trajectory of activating the servers as a function of $R_{\min}$. Second, designing waiting policies for the average AoI case. Third, considering different models for the servers including unresponsiveness, which requires employing robust PIR schemes as in \cite{JafarColluding,Staircase_PIR}. Fourth, proposing explicit PIR schemes to avoid the extra AoI penalty due to time sharing for the average AoI case. Finally, treating the message size $L$ as a control variable by which the user may receive shorter messages at a lower resolution versus awaiting fixed size ones.

\newpage
\bibliographystyle{unsrt}
\bibliography{reference}

\end{document}